\documentclass[%
 reprint,
 amsmath,amssymb,
 aps,
 prl,
 twocolumn,
 superscriptaddress
]{revtex4-2}
\setcitestyle{numbers,sort&compress}

\usepackage{graphicx}
\usepackage{dcolumn}
\usepackage{bm}
\usepackage{hyperref}
\usepackage{braket}
\usepackage{xcolor}
\usepackage{tikz}
\usepackage{ulem}
\usepackage{mhchem}
\usetikzlibrary{quantikz2}
\usetikzlibrary{arrows.meta, positioning}
\usepackage{subcaption}
\usepackage{multirow}
\usepackage{makecell}
\usepackage{amsthm}
\usepackage{soul}
\usepackage{float}
\usepackage{appendix}
\usepackage{algorithm}
\usepackage{algpseudocode}
\usepackage{ragged2e} 
\usepackage[table,xcdraw]{xcolor}

\begin{document}

\newcommand{\bea}{\begin{eqnarray}}
\newcommand{\eea}{\end{eqnarray}}
\newcommand{\EQ}[1]{Equation~(\ref{#1})} %
\newcommand{\eq}[1]{Eq.~(\ref{#1})} %
\newcommand{\eqs}[1]{Eqs.~(\ref{#1})} %
\newcommand{\fig}[1]{Fig.~\ref{#1}} %
\newcommand{\figs}[1]{Figs.~\ref{#1}} %
\newcommand{\CR}[1]{\hat a^{\dagger}_{#1}}
\newcommand{\AN}[1]{\hat a_{#1}}

\newcommand{\LA}[1]{\mathfrak{#1}}
\newcommand{\HG}{\hat G}

\newcommand{\BC}[1]{\hat \beta^{\dagger}_{#1}}
\newcommand{\BA}[1]{\hat \beta_{#1}}
\newcommand{\EU}[1]{\hat E^{#1}}
\newcommand{\ED}[1]{\hat E_{#1}}
\newcommand{\EE}[2]{\hat E_{#2}^{#1}}
\newcommand{\EX}[2]{\mathcal{E}_{#2}^{#1}}
\newcommand{\GP}[1]{\hat \gamma_{#1}}
\newcommand{\kh}{\hat \kappa}
\newcommand{\OO}{\hat O}
\newcommand{\HC}{\hat C}
\newcommand{\HA}[1]{\hat A_{#1}}


\newcommand{\bfr}{\mathbf{r}} %
\newcommand{\E}{\textrm{e}} %
\newcommand{\I}{\mathrm{i}\mkern1mu} %
\newcommand{\RR}{\mathbf{R}} %
\newcommand{\HP}{\hat P} %
\newcommand{\rr}{\mathbf{r}} %
\newenvironment{NB}{\color{red}NB: }{\ignorespacesafterend}  %
\newcommand{\HH}{\hat H} %
\newcommand{\HU}{\hat U} %
\newcommand{\hz}{\hat z} %

\newcommand{\AFI}[1]{\textcolor{blue}{[AFI: #1]}}

\preprint{APS/123-QED}

\title{Quantum Gambling: Best-Arm Strategies for Generator Selection in Adaptive Variational Algorithms}

\author{Rick Huang}
\affiliation{Chemical Physics Theory Group, Department of Chemistry, University of Toronto, Toronto, Ontario M5S 3H6, Canada}
\author{Artur F. Izmaylov}%
\email{artur.izmaylov@utoronto.ca}
\affiliation{Chemical Physics Theory Group, Department of Chemistry, University of Toronto, Toronto, Ontario M5S 3H6, Canada}
\affiliation{Department of Physical and Environmental Sciences, University of Toronto Scarborough, Toronto, Ontario M1C 1A4, Canada}

\date{\today}
\begin{abstract}
Adaptive variational algorithms suffer from prohibitively high measurement costs during the generator selection step, since energy gradients must be estimated for a large operator pool. This scaling bottleneck limits their applicability to larger molecular systems on near-term quantum devices.
We address this challenge by reformulating generator selection as a Best Arm Identification (BAI) problem, where the goal is to identify the generator with the largest energy gradient using as few measurements as possible. To solve it, we employ the Successive Elimination algorithm, which adaptively allocates measurements and discards unpromising candidates early.
Numerical experiments on molecular systems demonstrate that this approach substantially reduces the number of measurements required while preserving ground-state energy accuracy. By cutting measurement overhead without sacrificing performance, our method makes adaptive variational algorithms more practical for near-term quantum simulations.

\end{abstract}

\maketitle

\section{Introduction}

Variational quantum algorithms (VQAs) are widely studied as practical approaches for quantum simulation on near-term devices \cite{Cerezo2021variational}, while also providing a pathway to fault-tolerant quantum computing by enabling efficient preparation of correlated initial states for long-term algorithms \cite{choi2024probing}. Among these, adaptive variational algorithms dynamically construct an ansatz by selecting and appending parametrized unitaries generated from an operator pool $\mathcal{A} = \{\hat{G}_i\}$ \cite{grimsley2019adaptive}. At iteration $k$, the  wavefunction is
\begin{equation}
    \ket{\psi_k} = \prod_{i=1}^{k} e^{\theta_i \hat{G}_i} \ket{\psi_0},
\end{equation}
where $\ket{\psi_0}$ is the Hartree--Fock reference state.
The key task at each step is to identify the generator that most effectively lowers the energy. This is commonly achieved by evaluating the magnitude of the energy gradient
\begin{equation}
    g_i = \bra{\psi_k} [\hat{H}, \hat{G}_i] \ket{\psi_k},
\end{equation}
for each $\hat{G}_i \in \mathcal{A}$, and selecting $\hat{G}_M = \arg\max_i |g_i|$
with the goal of accelerating convergence toward the ground state.  

The challenge is that evaluating $g_i$ for each candidate generator requires decomposing commutators into measurable fragments, leading to a measurement cost that can scale as steeply as $\mathcal{O}(N^8)$ with the number of spin-orbitals. This scaling makes gradient estimation the dominant bottleneck in adaptive algorithms and a major obstacle to simulating chemically relevant molecular systems.  

A variety of approaches have been proposed to reduce this measurement overhead. One direction focuses on reducing the size of the operator pool. For instance, qubit-based operator pools of size $2N - 2$ have been shown to be expressive enough to represent the full Hilbert space \cite{tang2021qubit}. Later refinements exploited molecular symmetries to construct similarly compact pools while preserving conserved quantum numbers such as particle number and spin projection \cite{Shkolnikov2023Symmetry}. While effective in principle, these reductions 
increase the risk of trapping the ansatz in local minima.  

Another line of work reformulates the evaluation of gradients in terms of reduced density matrices (RDMs). For single and double excitation operators, gradients can be expressed without requiring more than three-body RDMs, and by approximating the latter in terms of lower-order RDMs, the measurement cost can be reduced from $\mathcal{O}(N^8)$ to $\mathcal{O}(N^4)$ \cite{Liu2021AdaptiveRDMSolver}. This reformulation takes advantage of the structure of the problem to dramatically reduce scaling while retaining accuracy in practice.  

Other strategies aim to improve the efficiency of generator selection itself. Instead of adding only the generator with the largest gradient at each iteration, one can include multiple generators whose gradients fall within a chosen threshold of the maximum. This approach accelerates convergence by requiring fewer iterations to construct an accurate ansatz, thereby reducing overall measurement demands \cite{Sapova2022, Lan2022AmplitudeReordering}. A complementary idea bundles qubit-based operators into a smaller number of measurement groups, reducing the scaling of gradient evaluation from $\mathcal{O}(N^8)$ to $\mathcal{O}(N^5)$ \cite{anastasiou2023measure}. These strategies show that smarter selection and grouping procedures can substantially reduce the depth of ansatz construction and the number of measurements needed per iteration.  

Finally, several methods exploit correlations between successive iterations to recycle measurement information. Since commutator decompositions often share measurable fragments with the Hamiltonian itself, it is possible to reuse measurement data from the VQE subroutine in subsequent gradient evaluations \cite{Ikhtiarudin2025,AIMADAPTVQE2022}. Combined with adaptive allocation rules, such as allocating shots proportional to the empirical variance of measurement groups, these techniques reduce unnecessary sampling compared to uniform allocation. Related approaches embed effective operators identified in smaller systems into larger Hilbert spaces to avoid re-estimating their effectiveness from scratch \cite{vandyke2024operator}.  

Despite this broad range of advances, all existing methods share a critical limitation: they require estimating the gradients of a large fraction of the operator pool to a fixed precision, even though most operators contribute negligibly to lowering the energy. As a result, measurement resources are wasted on accurately characterizing unpromising candidates that are unlikely ever to be selected. This inefficiency highlights a conceptual gap in the current literature: while existing methods reduce scaling, they do not fundamentally change the need to measure every candidate to essentially the same level of accuracy. 
In this work, we take a different perspective by viewing generator selection as a resource allocation problem under uncertainty. At each iteration, the adaptive algorithm must distribute a finite measurement budget across many candidate generators whose true gradients are unknown. This scenario is closely related to the \textit{Best Arm Identification} (BAI) problem \cite{EvenDar2006Action,Audibert2010BestArm,Bubeck2012Bandits}, in which one repeatedly samples from different options (``arms'') to determine which has the largest mean reward.  

This analogy maps naturally onto adaptive variational algorithms: each generator corresponds to an arm, and its ``reward’’ is the energy gradient signaling its effectiveness in lowering the system’s energy. We apply the \textit{Successive Elimination} algorithm \cite{EvenDar2006Action}, a well-established BAI solver, to adaptively allocate measurements and discard unpromising candidates early. By focusing resources on the few generators most likely to drive convergence, this approach avoids the uniform precision requirement of earlier methods.  

\section{Methods}
\label{sec:methods}

\begin{figure}[h!]
\centering
\begin{tikzpicture}[node distance=1.5cm,>=stealth,thick]

\tikzstyle{block} = [rectangle, rounded corners, draw=black, text centered, text width=6.5cm, minimum height=1cm, fill=blue!5]
\tikzstyle{line} = [draw, -{Latex[length=3mm]}]

\node[block] (init) {Start from reference state $\ket{\psi_0}$ (Hartree--Fock)};
\node[block, below of=init] (ansatz) {Current ansatz $\ket{\psi_k} = \prod_{i=1}^k e^{\theta_i \hat{G}_i} \ket{\psi_0}$};
\node[block, below of=ansatz] (commutator) {Decompose commutators to measurable fragments: $[\hat{H}, \hat{G}_i] = \sum_n \hat{A}_n^{(i)}$};
\node[block, below of=commutator] (clt) {Estimate fragment expectation values $\braket{\hat{A}_n^{(i)}}$};
\node[block, below of=clt] (se) {Apply \textit{Successive Elimination}: adaptively estimate gradients, eliminate weak candidates, select best $\hat{G}_M$};
\node[block, below of=se] (update) {Update ansatz: $\ket{\psi_{k+1}} = e^{\theta_{k+1}\hat{G}_M}\ket{\psi_k}$};
\node[block, below of=update] (vqe) {Run VQE global re-optimization of parameters $\{\theta_i\}$ (L-BFGS-B)};
\node[block, below of=vqe] (loop) {Repeat until convergence or maximum iterations};

\path[line] (init) -- (ansatz);
\path[line] (ansatz) -- (commutator);
\path[line] (commutator) -- (clt);
\path[line] (clt) -- (se);
\path[line] (se) -- (update);
\path[line] (update) -- (vqe);
\path[line] (vqe) -- (loop);

\end{tikzpicture}
\caption{Flowchart of the adaptive variational scheme with successive elimination for generator selection. The procedure alternates between ansatz construction, gradient estimation, elimination-based selection, and parameter re-optimization.}
\label{fig:flowchart}
\end{figure}
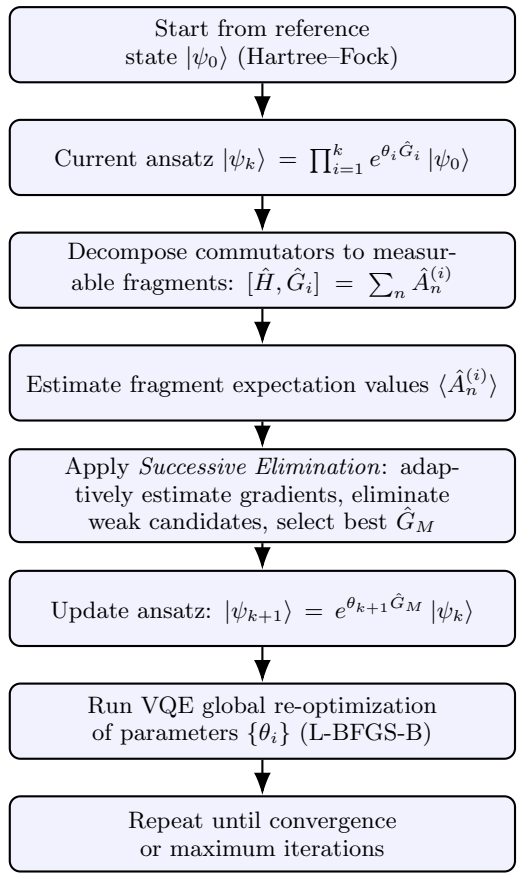

A flow-chart of a typical adaptive variational algorithm is given by Figure~\ref{fig:flowchart}, and in what follows we elaborate on each part, starting with the main contribution of this work. 

\subsection{Successive Elimination for Generator Selection}
Here we illustrate the search for the largest gradient $\{g_i\}$ using 
the \textit{Successive Elimination} (SE) algorithm to minimize the number of measurements. SE adaptively allocates measurements across rounds and progressively eliminates candidates with small gradients, concentrating sampling effort on promising generators. 

The SE algorithm runs a loop of several rounds enumerated by $r$. 
At round $r$, let $A_r \subseteq \mathcal{A}$ denote the active set of generators (with $A_0 = \mathcal{A}$). The procedure is as follows:
\begin{enumerate}
    \item \textbf{Initialization:} Begin with the state $\ket{\psi_k}$ obtained from the last VQE optimization.  
    \item \textbf{Adaptive Measurements:} For each $\hat{G}_i \in A_r$, estimate $g_i$ with precision $\epsilon_r = c_r \cdot \epsilon$ ($c_r \geq 1$).  
    \item \textbf{Gradient Estimation:} Compute $|g_i|$ by summing estimated expectation values of its measurable fragments.  
    \item \textbf{Candidate Elimination:} Let $M = \max_i |g_i|$ within $A_r$. Eliminate all generators $\hat{G}_i$ such that
    \begin{equation}
        |g_i| + R_r < M - R_r,
    \end{equation}
    where $R_r = d_r \cdot \epsilon_r$ with $d_r$ a constant.  
    \item \textbf{Termination:} Continue until either only one candidate remains or the maximum number of rounds $L$ is reached. In the latter case, the generator with the largest gradient is chosen.
\end{enumerate}
In the final round ($r=L$), we set $c_L = 1$ so that the selected gradient is estimated to the target accuracy $\epsilon$.

\subsection{Gradient Estimation via Fragmentation and Sampling}
To implement successive elimination, we must estimate gradients from quantum measurements. For each generator $\hat{G}_i \in \mathcal{A}$, the commutator decomposes into a sum of measurable fragments,
\begin{equation}\label{eq:MF}
    [\hat{H}, \hat{G}_i] = \sum_n \hat{A}_n^{(i)},
\end{equation}
which yields
\begin{equation}
    g_i = \sum_n \braket{\hat{A}_n^{(i)}}.
\end{equation}
A variety of measurable fragmentations can be used in \eq{eq:MF} \cite{patel2025quantum}. In this work, we employ one of the simplest: 
qubit-wise commuting (QWC) fragmentation approach with the sorted insertion 
(SI) grouping strategy\cite{yen2023deterministic}. 

Each fragment $\hat{A}_n^{(i)}$ is measured through repeated sampling. By the Central Limit Theorem, the distribution of the empirical mean converges to a normal distribution with mean
\begin{equation}
    \mathbb{E}[\hat{A}_n^{(i)}] = \braket{\hat{A}_n^{(i)}},
\end{equation}
and with variance given by
\begin{equation}
    Var(\hat{A}_n^{(i)}) = \braket{(\hat{A}_n^{(i)})^2} - \braket{\hat{A}_n^{(i)}}^2,
\end{equation}
where all expectation values are taken with the wavefunction obtained after the most recent VQE parameter reoptimization. These quantities define the normal distribution $P_n^{(i)}$ used to model measurement statistics.  

In the na\"ive strategy, each fragment requires
\begin{equation}
    M_n(\epsilon) = \frac{Var(\hat{A}_n^{(i)})}{\epsilon^2}
\end{equation}
measurements to achieve precision $\epsilon$. Successive elimination improves upon this by allocating fewer measurements to fragments of weak generator candidates, thereby reducing overall cost.

\subsection{Operator Pools}
The choice of operator pool influences both expressibility and measurement cost. To demonstrate the generality of our approach, we benchmark across several widely used pools.

The \textbf{UCCSD pool} \cite{grimsley2019adaptive} consists of fermionic single and double excitation operators that preserve spin symmetries:
\begin{equation}
    \{a^\dagger_a a_i - a^\dagger_i a_a,\;
      a^\dagger_a a^\dagger_b a_i a_j - a^\dagger_j a^\dagger_i a_b a_a \},
\end{equation}
where $a,b$ denote virtual orbitals and $i,j$ occupied orbitals of the Hartree--Fock state.

The \textbf{Qubit pool} \cite{tang2021qubit} is obtained by applying the Jordan--Wigner transformation to UCCSD operators and removing Pauli-$Z$ operators:
\begin{equation}
    \{iX_pY_q,\; iX_pY_qY_rY_s,\; iY_pX_qX_rX_s\},
\end{equation}
where $p,q,r,s$ are qubit indices.

The \textbf{Qubit excitation pool} \cite{yordanov2021qubit} employs qubit creation and annihilation operators,
\begin{equation}
    \hat{Q}^\dagger_p = \frac{X_p - iY_p}{2}, \quad
    \hat{Q}_p = \frac{X_p + iY_p}{2},
\end{equation}
to define
\begin{equation}
    \{\hat{Q}^\dagger_p \hat{Q}_q - \hat{Q}_q^\dagger \hat{Q}_p,\;
      \hat{Q}^\dagger_p \hat{Q}^\dagger_q \hat{Q}_r \hat{Q}_s
      - \hat{Q}_s^\dagger \hat{Q}_r^\dagger \hat{Q}_q \hat{Q}_p\}.
\end{equation}

By testing across these pools, we verify that successive elimination improves measurement efficiency regardless of pool structure.

\subsection{VQE Parameter Optimization}
Once a generator is selected, the ansatz is updated and parameters are re-optimized using a VQE subroutine. We adopt global re-optimization, where all parameters are optimized simultaneously, to maintain consistency and prevent parameter drift. The cost function is
\begin{equation}
    f(\Vec{\theta}) = \bra{\psi_0}
    \prod_{i=k+1}^{1} e^{\hat{G}_i \theta_i}
    \hat{H}
    \prod_{i=1}^{k+1} e^{-\hat{G}_i \theta_i}
    \ket{\psi_0},
\end{equation}
with $\Vec{\theta} = (\theta_1, \dots, \theta_{k+1})$. 
Optimization is performed using the exact state-vector representation of the wavefunction with the L-BFGS-B algorithm implemented in \texttt{scipy.optimize.minimize}. Convergence is determined when the gradient norm falls below a threshold or when the maximum number of iterations is reached.

\section{Results and Discussion}

    \begin{figure*}[htbp]
        \centering
        \begin{subfigure}[b]{0.32\textwidth}
            \includegraphics[width=1.0\textwidth]{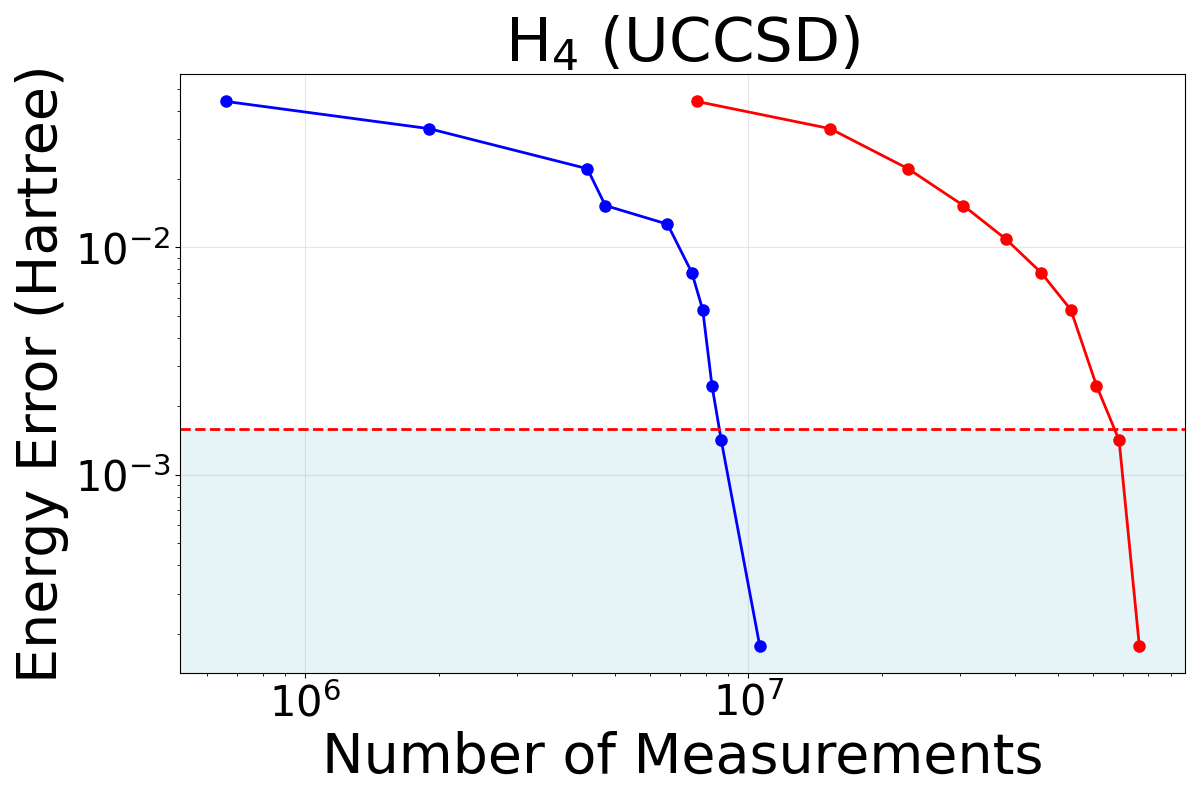}
            \label{fig:h4_uccsd}
        \end{subfigure}
        \begin{subfigure}[b]{0.32\textwidth}
            \includegraphics[width=1.0\textwidth]{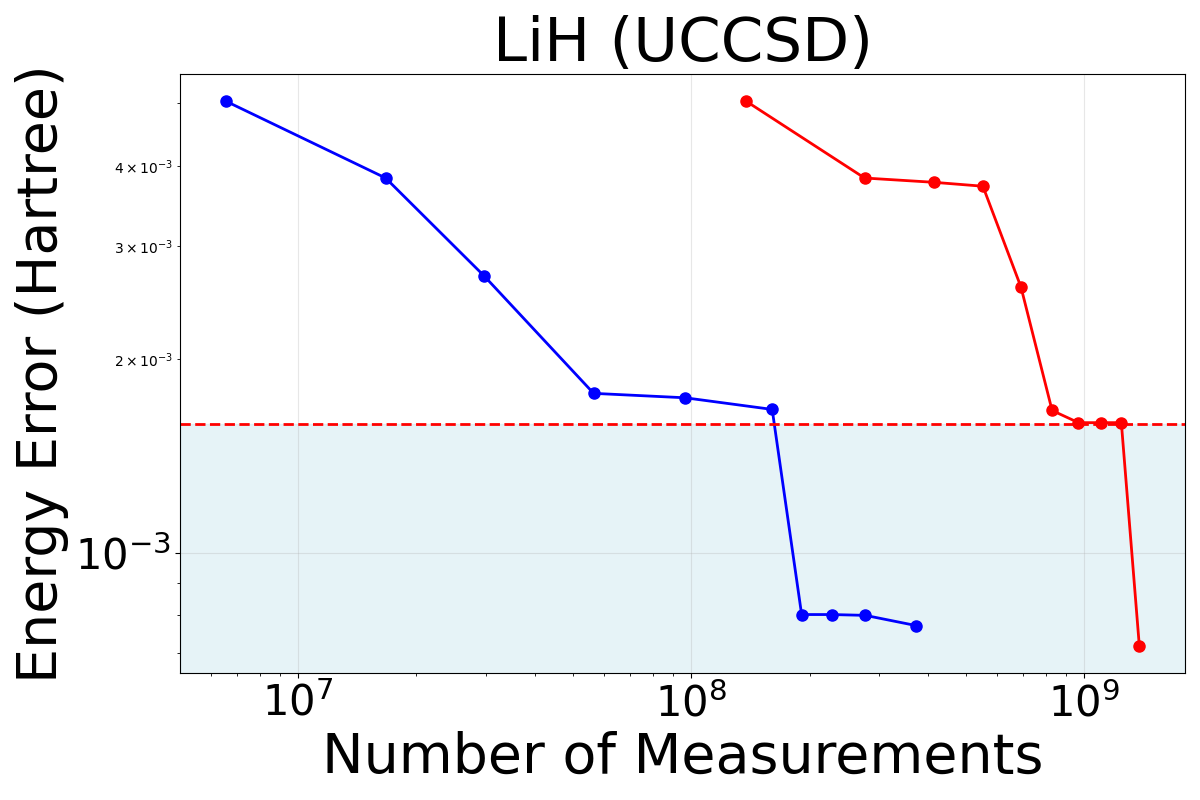}
            \label{fig:single-}
        \end{subfigure}
        \begin{subfigure}[b]{0.32\textwidth}
           \includegraphics[width=1.0\textwidth]{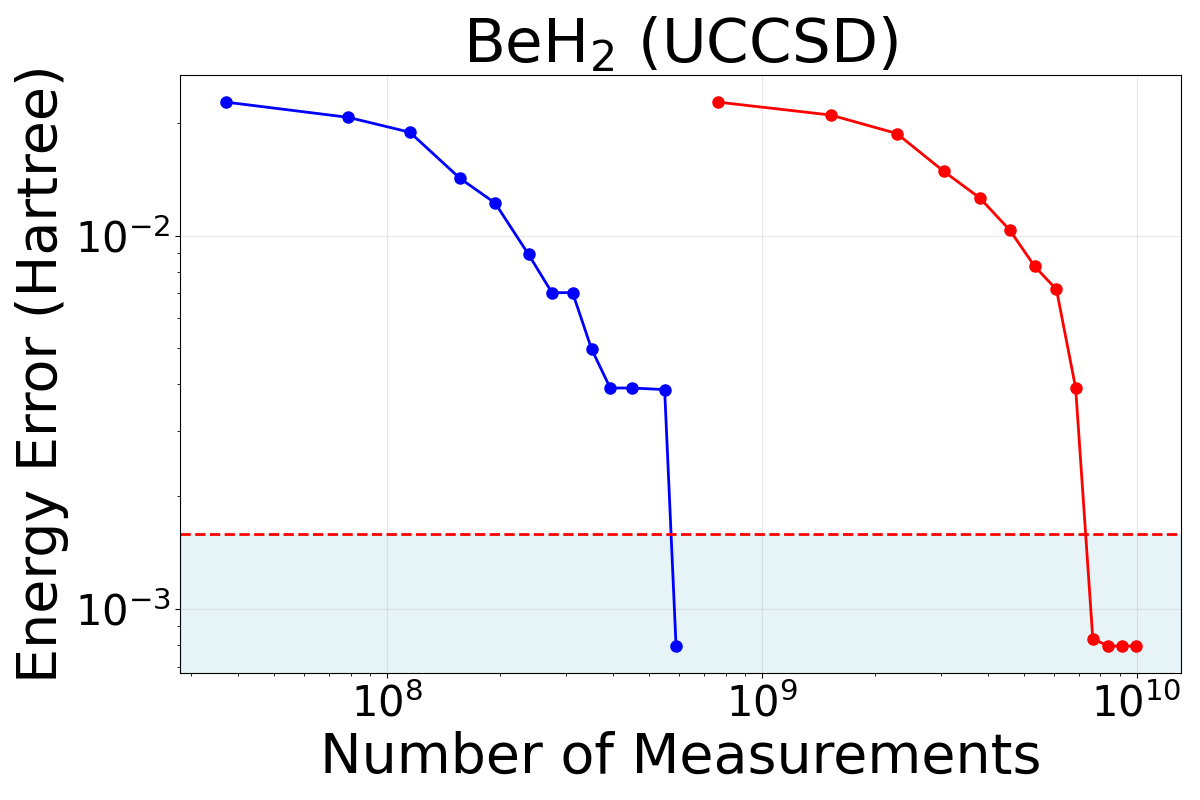}
            \label{fig:single-image}
        \end{subfigure}
        
        \vspace{1em} 
    
        \begin{subfigure}[b]{0.32\textwidth}
           \includegraphics[width=1.0\textwidth]{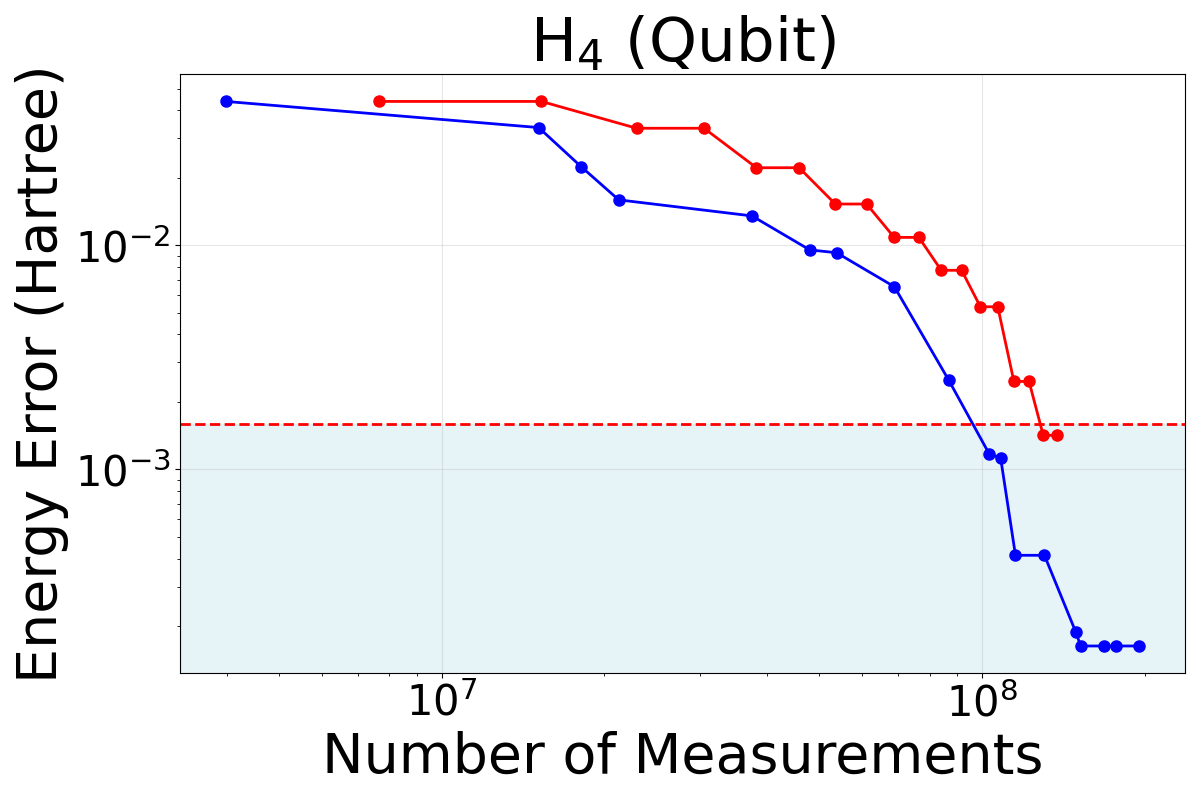}
            \label{fig:single-image}
        \end{subfigure}
        \begin{subfigure}[b]{0.32\textwidth}
           \includegraphics[width=1.0\textwidth]{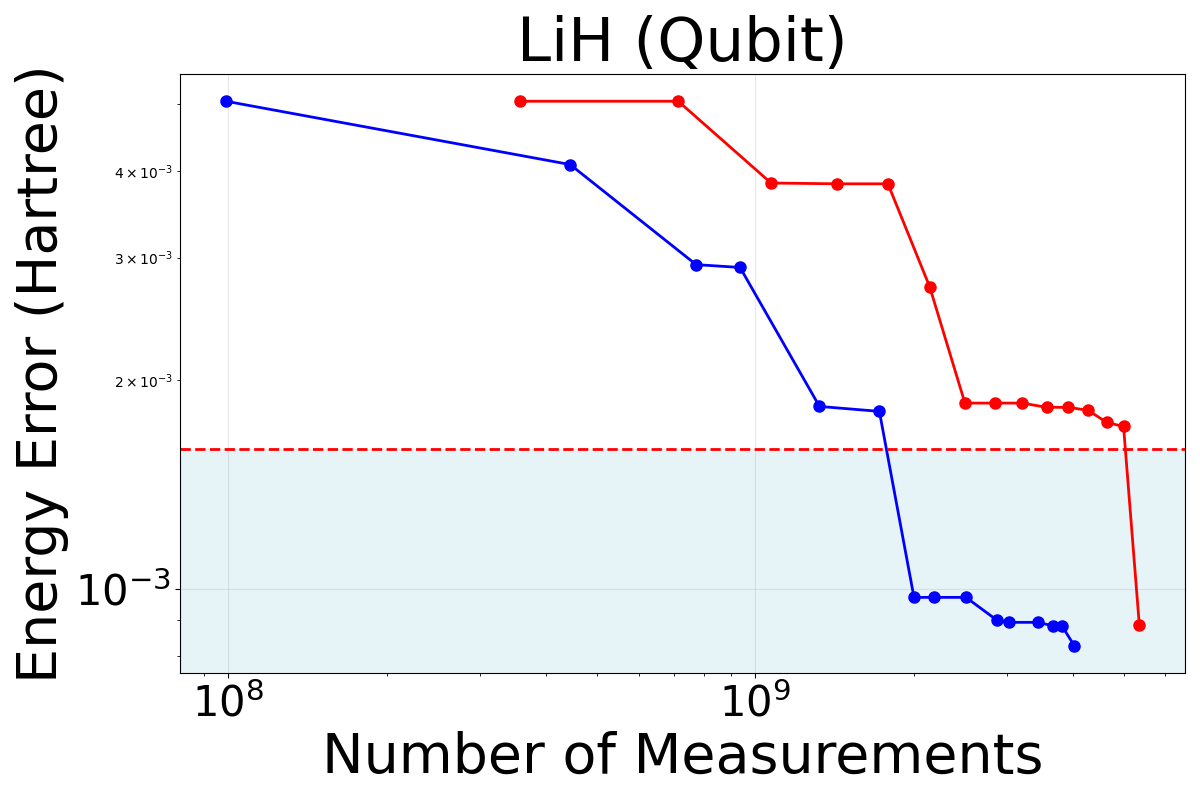}
            \label{fig:single-image}
        \end{subfigure}
        \begin{subfigure}[b]{0.32\textwidth}
            \includegraphics[width=1.0\textwidth]{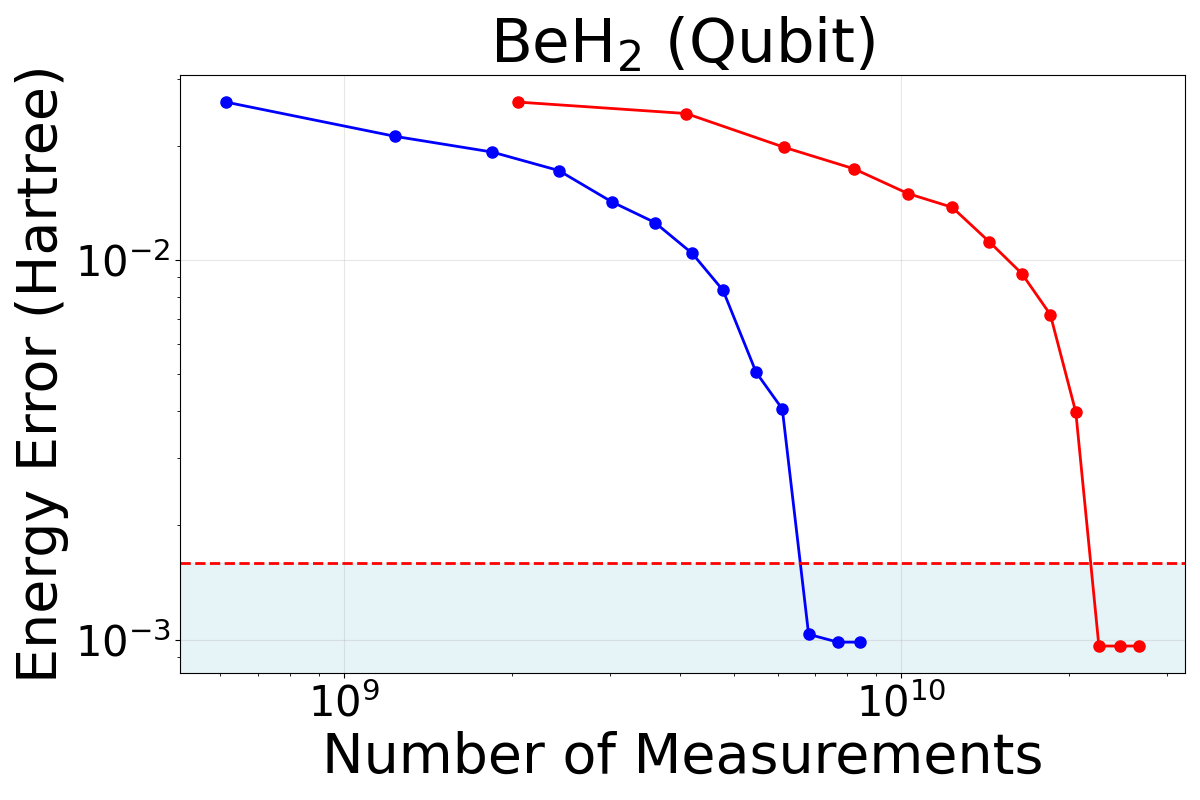}
            \label{fig:single-image}
        \end{subfigure}

        \vspace{1em} 
    
        \begin{subfigure}[b]{0.32\textwidth}
           \includegraphics[width=1.0\textwidth]{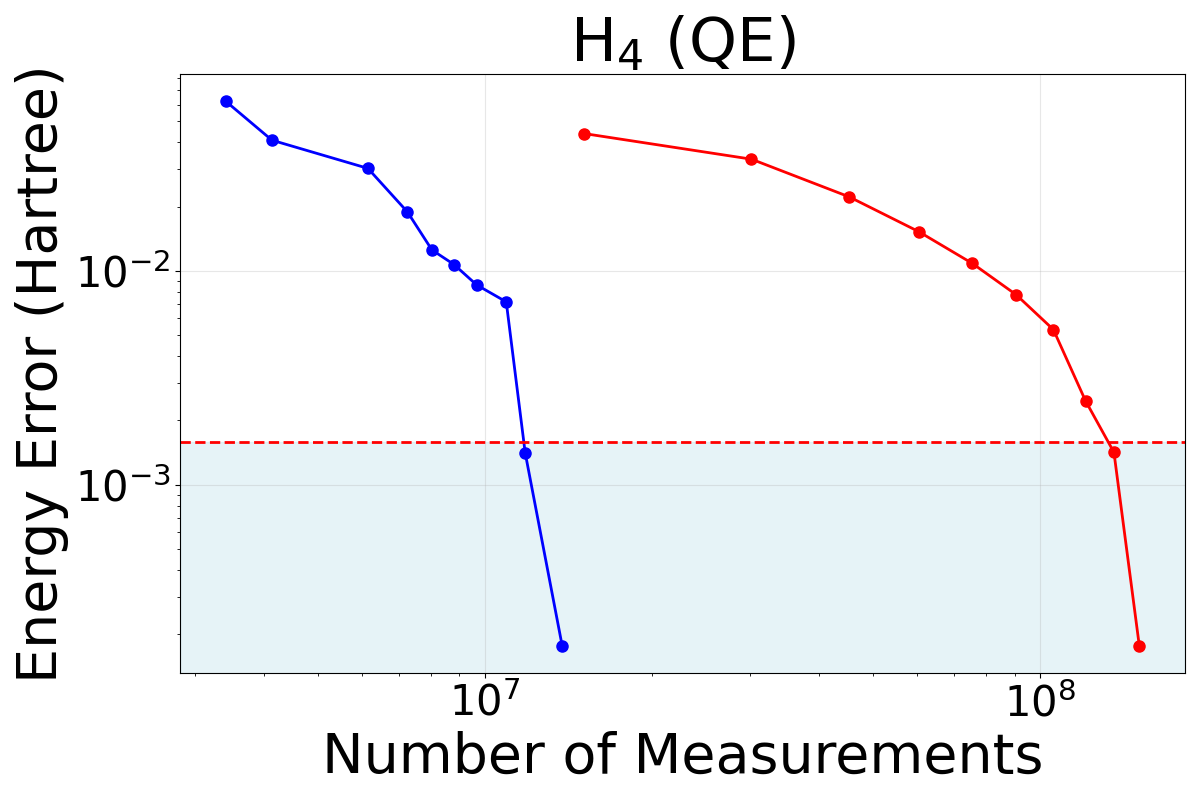}
            \label{fig:qubit_excitation}
        \end{subfigure}
        \begin{subfigure}[b]{0.32\textwidth}
           \includegraphics[width=1.0\textwidth]{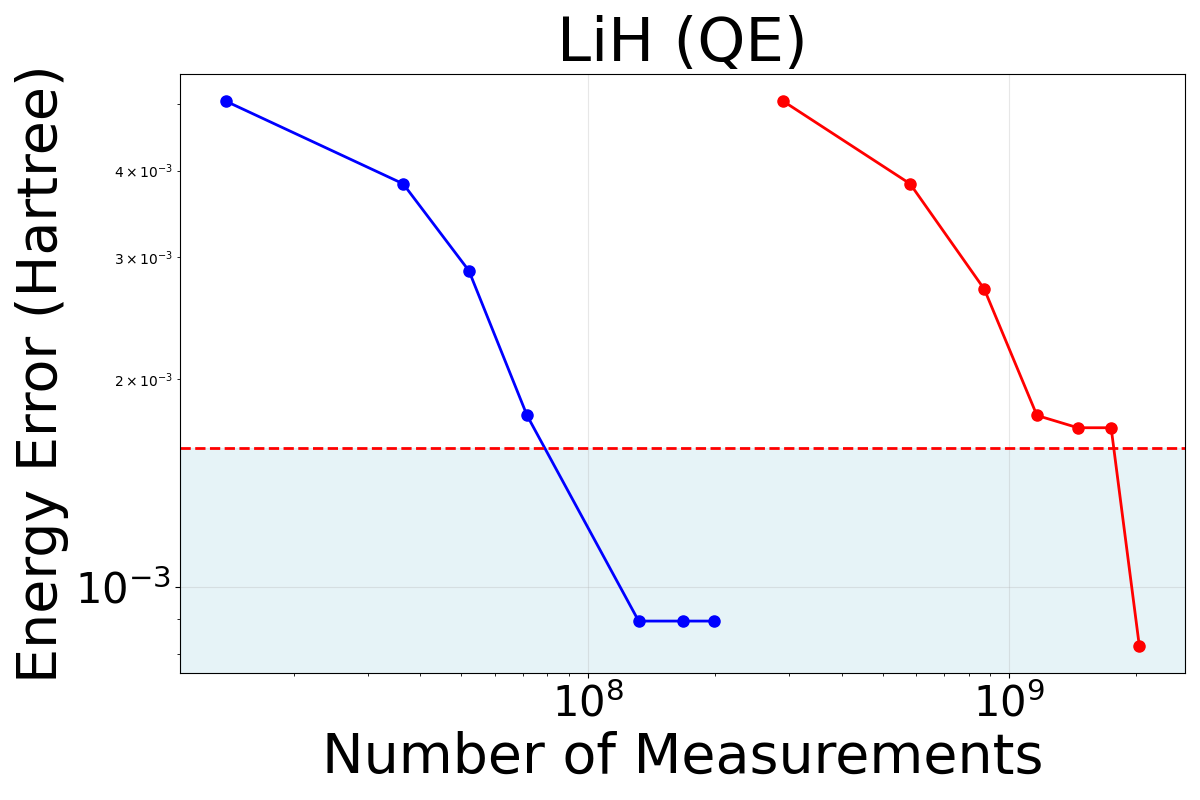}
            \label{fig:single-image}
        \end{subfigure}
        \begin{subfigure}[b]{0.32\textwidth}
            \includegraphics[width=1.0\textwidth]{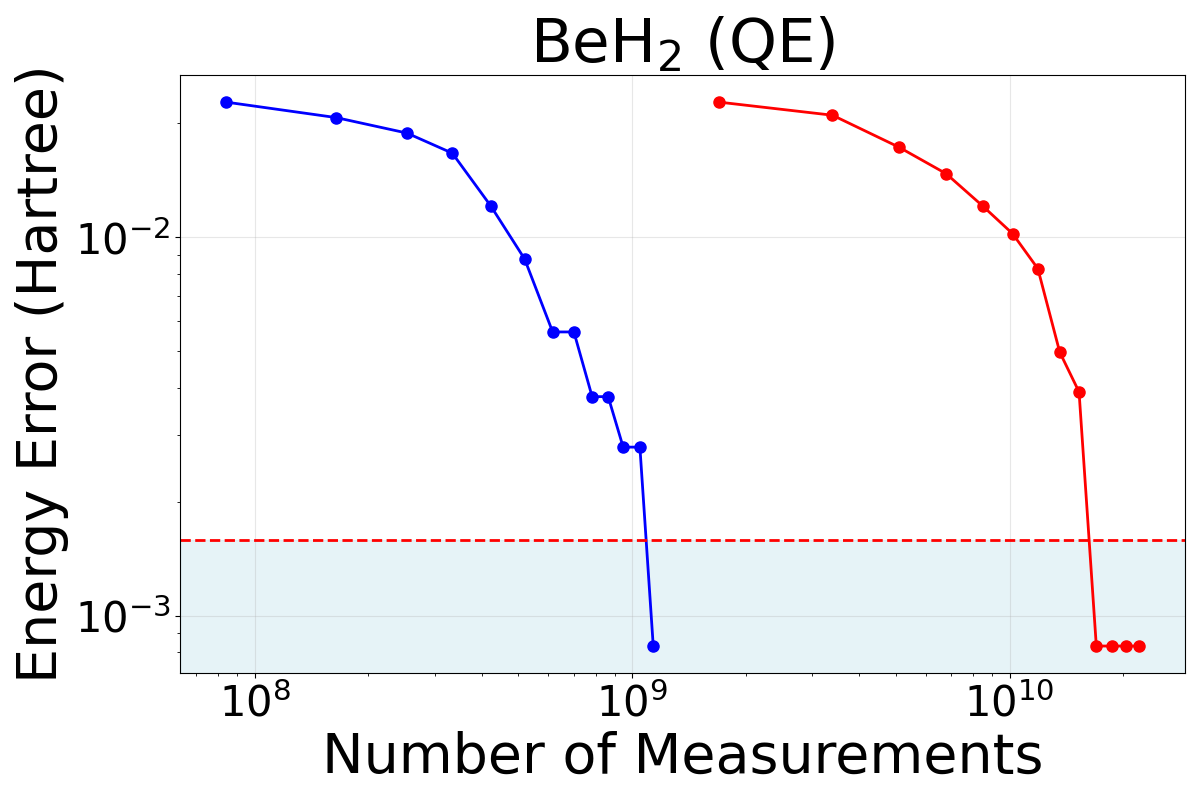}
            \label{fig:single-image}
        \end{subfigure}
    
        \caption{
        Energy error (in Hartree) versus cumulative number of measurements for single ADAPT-VQE generator selection process across three molecular systems (H$_4$, LiH, BeH$_2$) and three operator pools (UCCSD, Qubit, Qubit-Excitation). 
    In each panel, the \textcolor{red}{red curve} corresponds to the na\"ive baseline in which all gradients are estimated to fixed precision 
    ($\epsilon = 1.0\times 10^{-3}$ for UCCSD and Qubit-Excitation pools, $\epsilon = 0.5\times 10^{-3}$ for the Qubit pool). 
    The \textcolor{blue}{blue curve} shows the proposed \textit{Successive Elimination} (SE) strategy, which adaptively allocates measurements during gradient evaluation. 
    The horizontal red line indicates chemical accuracy ($1.59\times 10^{-3}$ Hartree). 
    Across all systems and pools, SE achieves convergence within chemical accuracy while requiring substantially fewer measurements compared to the na\"ive approach.}
        \label{fig:result plot}
    \end{figure*}

We benchmarked the proposed \textit{Successive Elimination} (SE) strategy against the na\"ive fixed-precision baseline for generator selection in adaptive variational algorithms. Our goal was to quantify the reduction in measurement cost while maintaining chemical accuracy in the computed ground-state energies.  

\subsection{Benchmark systems and setup}

We considered three molecular systems of increasing size and electronic complexity—H$_4$, LiH, and BeH$_2$—all in the STO-3G basis set. 
Hamiltonians were generated with \texttt{PySCF}, and fermion-to-qubit mappings were carried out using the Jordan–Wigner transformation. For all three systems, we used a linear geometry with a bond distance of 1 \AA between adjacent atoms.

We tested three widely used operator pools: UCCSD \cite{grimsley2019adaptive}, Qubit\cite{tang2021qubit}, and Qubit-Excitation (QE)\cite{yordanov2021qubit}. 
The Hartree--Fock wavefunction was used as the reference in all cases.  
Within each ADAPT iteration, parameters were globally re-optimized using L-BFGS-B, as described in Sec.~\ref{sec:methods}.

For the baseline method, each gradient was estimated with a target precision of $\epsilon = 1.0 \times 10^{-3}$ for the UCCSD and QE pools, and $\epsilon = 0.5 \times 10^{-3}$ for the Qubit pool.
For the SE-based method, the per-round precision was defined as $\epsilon_r^\prime = (5 - 2r/5)\epsilon$ for the UCCSD and QE pools, and $\epsilon_r^\prime = (2 - r/10)\epsilon$ for the Qubit pool, with a maximum of 10 rounds. 
The confidence radius was set to $R_r=8\epsilon_r^\prime$. 
These hyperparameters were tuned based on numerical simulation results, as they consistently produced a substantial reduction in measurement cost.

\subsection{Measurement savings}

Figure~\ref{fig:result plot} compares the convergence of the energy error, relative to the exact ground-state energy, as a function of the cumulative number of measurements. Results 
are shown for both the na\"ive baseline and the SE-based strategy during generator selection. Across all systems and operator pools, the na\"ive baseline shows a steep initial rise in measurement cost because every generator gradient is estimated to high precision from the outset. By contrast, the SE approach grows more slowly in measurement cost, since weak candidates are discarded early and subsequent rounds concentrate measurements on only a few surviving operators. This leads to a ``step-like" progression in which the error reduction is punctuated by phases of low-cost elimination. Importantly, both methods converge to the same final energy within chemical accuracy, confirming that the elimination of suboptimal candidates does not compromise the ansatz construction. The figure also highlights a trade-off: in some cases SE requires more ADAPT iterations, since discarding near-optimal operators can delay the incorporation of certain important excitations. However, the additional VQE steps incur only marginal overhead compared to the substantial savings in gradient evaluation. 
This is because the measurement cost of VQE parameter reoptimization grows more slowly with system size, whereas generator selection requires an increasing number of expectation-value evaluations as the operator pool expands. Overall, the measurement budget is significantly more favorable for SE.
Table~\ref{reduction table} reports the percentage reduction in the total number of measurements required for generator selection to reach chemical accuracy.  

\begin{table}[htbp]
    \caption{Percentage reduction in the total number of measurements required for generator selection to reach chemical accuracy in the ADAPT-VQE algorithm when using Successive Elimination, relative to the naïve approach. Target precisions were $\epsilon = 1.0 \times 10^{-3}$ for UCCSD and QE pools, and $\epsilon = 0.5 \times 10^{-3}$ for the Qubit pool.}
    \centering
    \Large
    \begin{tabular}{c | c |c| c}
        \hline
        Pool & H$_4$ & LiH & BeH$_2$ \\ \hline
        UCCSD & $93.0$ & $80.8$ & $90.0$  \\ \hline
        Qubit & $69.4$ &  $70.9$ & $71.9$ \\ \hline
        QE & $92.2$ & $84.5$ & $89.8$ \\ \hline
    \end{tabular}
    \label{reduction table}
\end{table}

Across all operator pools and molecules, SE yields substantial savings in measurement cost, ranging from $\sim 70\%$ for the Qubit pool to over $90\%$ for UCCSD and QE pools. 
The larger savings in UCCSD and QE pools are due to their larger operator sets and higher fragment variances, which make early elimination of weak candidates especially beneficial.
By contrast, the Qubit pool is already compact, so the scope for improvement is smaller, though savings remain significant.

System size also plays a role: while the smallest system, H$_4$, shows the largest percentage reduction, both LiH and BeH$_2$ still achieve over $70\%$ savings. This suggests that the benefits of SE persist with increasing system size and may even scale favorably as operator pools grow.  

\subsection{Trade-offs and convergence behavior}

An important observation is that SE can increase the number of ADAPT iterations compared to the na\"ive baseline, since occasional elimination of near-optimal generators may delay convergence. 
However, the overall measurement cost remains much lower because evaluating the Hamiltonian in the VQE subroutine is considerably cheaper than computing all gradients in the generator pool. 
Moreover, as shown in Fig.~\ref{fig:result plot}, ground-state energies obtained with SE remain within chemical accuracy, confirming that the adaptive elimination process does not compromise the final result.  

In practice, the measurement efficiency of SE could be further improved by refining the precision schedule $\epsilon_r$ and confidence radius $R_r$, or by integrating with variance-aware measurement allocation strategies.  
These optimizations represent promising directions for tailoring BAI-inspired selection to quantum chemistry applications.  

\subsection{Implications}

The results demonstrate that reframing generator selection as a BAI problem offers a principled route to scalable resource allocation in adaptive variational algorithms. 
Rather than treating all operators equally, SE directs sampling effort toward the most promising candidates and avoids wasting measurements on negligible gradients. 
The consistent $60$--$90\%$ savings observed here suggest that this strategy can significantly lower the experimental overhead of adaptive methods, moving them closer to practical deployment on near-term quantum devices and providing a foundation for efficient state preparation in future fault-tolerant simulations.  

\section{Conclusion}
In this work, we have addressed the high measurement cost of generator selection in adaptive variational quantum algorithms (AVQAs) by reformulating the problem as an instance of \textit{Best Arm Identification} (BAI). Within this framework, generator selection is viewed as an adaptive resource allocation task under uncertainty, and we demonstrated a proof-of-concept implementation using the classical \textit{Successive Elimination} (SE) algorithm. Our numerical benchmarks on H$_4$, LiH, and BeH$_2$ across several widely used operator pools show that this strategy can reduce the measurement overhead for gradient estimation by $60\%$--$92\%$, while preserving chemical accuracy.  

Because the SE-based approach targets measurement effort only toward promising candidates, it offers a conceptually different and more efficient path than existing strategies that estimate all gradients to uniform precision. Moreover, the method is compatible with previous advances in pool design \cite{tang2021qubit, Shkolnikov2023Symmetry} and measurement reuse \cite{Ikhtiarudin2025, AIMADAPTVQE2022}, and can therefore be integrated to achieve further cost reductions. Importantly, the reformulation as a BAI problem is not tied to any specific operator pool or VQE optimization details, suggesting that it may generalize broadly within adaptive variational schemes.  

Looking ahead, future work could explore alternative BAI solvers such as \textit{Track-and-Stop} \cite{pmlr-v49-garivier16a} or Bayesian approaches \cite{russo2020simple}, which may yield additional savings or improved robustness. More generally, the ``quantum gambling'' perspective developed here highlights a scalable principle for making adaptive algorithms more resource-efficient, and provides a pathway toward practical quantum simulations both on near-term devices and as state-preparation tools for fault-tolerant quantum computing.

\section*{Acknowledgments}
Authors thank Tzu-Ching Yen for useful discussions. This research has received funding from the research project entitled “Quantum Software Consortium: Exploring Distributed Quantum Solutions for Canada” (QSC).  QSC is financed by the National Sciences and Engineering Research Council of Canada (NSERC) Alliance Consortia Quantum program under grant number ALLRP587590-23.

\appendix

\bibliographystyle{apsrev4-2}
\bibliography{library}

\begin{thebibliography}{20}%
\makeatletter
\providecommand \@ifxundefined [1]{%
 \@ifx{#1\undefined}
}%
\providecommand \@ifnum [1]{%
 \ifnum #1\expandafter \@firstoftwo
 \else \expandafter \@secondoftwo
 \fi
}%
\providecommand \@ifx [1]{%
 \ifx #1\expandafter \@firstoftwo
 \else \expandafter \@secondoftwo
 \fi
}%
\providecommand \natexlab [1]{#1}%
\providecommand \enquote  [1]{``#1''}%
\providecommand \bibnamefont  [1]{#1}%
\providecommand \bibfnamefont [1]{#1}%
\providecommand \citenamefont [1]{#1}%
\providecommand \href@noop [0]{\@secondoftwo}%
\providecommand \href [0]{\begingroup \@sanitize@url \@href}%
\providecommand \@href[1]{\@@startlink{#1}\@@href}%
\providecommand \@@href[1]{\endgroup#1\@@endlink}%
\providecommand \@sanitize@url [0]{\catcode `\\12\catcode `\$12\catcode `\&12\catcode `\#12\catcode `\^12\catcode `\_12\catcode `\%12\relax}%
\providecommand \@@startlink[1]{}%
\providecommand \@@endlink[0]{}%
\providecommand \url  [0]{\begingroup\@sanitize@url \@url }%
\providecommand \@url [1]{\endgroup\@href {#1}{\urlprefix }}%
\providecommand \urlprefix  [0]{URL }%
\providecommand \Eprint [0]{\href }%
\providecommand \doibase [0]{https://doi.org/}%
\providecommand \selectlanguage [0]{\@gobble}%
\providecommand \bibinfo  [0]{\@secondoftwo}%
\providecommand \bibfield  [0]{\@secondoftwo}%
\providecommand \translation [1]{[#1]}%
\providecommand \BibitemOpen [0]{}%
\providecommand \bibitemStop [0]{}%
\providecommand \bibitemNoStop [0]{.\EOS\space}%
\providecommand \EOS [0]{\spacefactor3000\relax}%
\providecommand \BibitemShut  [1]{\csname bibitem#1\endcsname}%
\let\auto@bib@innerbib\@empty
\bibitem [{\citenamefont {Cerezo}\ \emph {et~al.}(2021)\citenamefont {Cerezo}, \citenamefont {Arrasmith}, \citenamefont {Babbush}, \citenamefont {Benjamin}, \citenamefont {Endo}, \citenamefont {Fujii}, \citenamefont {McClean}, \citenamefont {Mitarai}, \citenamefont {Yuan}, \citenamefont {Cincio},\ and\ \citenamefont {Coles}}]{Cerezo2021variational}%
  \BibitemOpen
  \bibfield  {author} {\bibinfo {author} {\bibfnamefont {M.}~\bibnamefont {Cerezo}}, \bibinfo {author} {\bibfnamefont {A.}~\bibnamefont {Arrasmith}}, \bibinfo {author} {\bibfnamefont {R.}~\bibnamefont {Babbush}}, \bibinfo {author} {\bibfnamefont {S.~C.}\ \bibnamefont {Benjamin}}, \bibinfo {author} {\bibfnamefont {S.}~\bibnamefont {Endo}}, \bibinfo {author} {\bibfnamefont {K.}~\bibnamefont {Fujii}}, \bibinfo {author} {\bibfnamefont {J.~R.}\ \bibnamefont {McClean}}, \bibinfo {author} {\bibfnamefont {K.}~\bibnamefont {Mitarai}}, \bibinfo {author} {\bibfnamefont {X.}~\bibnamefont {Yuan}}, \bibinfo {author} {\bibfnamefont {L.}~\bibnamefont {Cincio}},\ and\ \bibinfo {author} {\bibfnamefont {P.~J.}\ \bibnamefont {Coles}},\ }\href {https://doi.org/10.1038/s42254-021-00348-9} {\bibfield  {journal} {\bibinfo  {journal} {Nature Reviews Physics}\ }\textbf {\bibinfo {volume} {3}},\ \bibinfo {pages} {625} (\bibinfo {year} {2021})}\BibitemShut {NoStop}%
\bibitem [{\citenamefont {Choi}\ \emph {et~al.}(2024)\citenamefont {Choi}, \citenamefont {Loaiza}, \citenamefont {Lang}, \citenamefont {Mart{\'\i}nez-Mart{\'\i}nez},\ and\ \citenamefont {Izmaylov}}]{choi2024probing}%
  \BibitemOpen
  \bibfield  {author} {\bibinfo {author} {\bibfnamefont {S.}~\bibnamefont {Choi}}, \bibinfo {author} {\bibfnamefont {I.}~\bibnamefont {Loaiza}}, \bibinfo {author} {\bibfnamefont {R.~A.}\ \bibnamefont {Lang}}, \bibinfo {author} {\bibfnamefont {L.~A.}\ \bibnamefont {Mart{\'\i}nez-Mart{\'\i}nez}},\ and\ \bibinfo {author} {\bibfnamefont {A.~F.}\ \bibnamefont {Izmaylov}},\ }\href@noop {} {\bibfield  {journal} {\bibinfo  {journal} {J. Chem. Theory Comput.}\ }\textbf {\bibinfo {volume} {20}},\ \bibinfo {pages} {5982} (\bibinfo {year} {2024})}\BibitemShut {NoStop}%
\bibitem [{\citenamefont {Grimsley}\ \emph {et~al.}(2019)\citenamefont {Grimsley}, \citenamefont {Economou}, \citenamefont {Barnes},\ and\ \citenamefont {Mayhall}}]{grimsley2019adaptive}%
  \BibitemOpen
  \bibfield  {author} {\bibinfo {author} {\bibfnamefont {H.~R.}\ \bibnamefont {Grimsley}}, \bibinfo {author} {\bibfnamefont {S.~E.}\ \bibnamefont {Economou}}, \bibinfo {author} {\bibfnamefont {E.}~\bibnamefont {Barnes}},\ and\ \bibinfo {author} {\bibfnamefont {N.~J.}\ \bibnamefont {Mayhall}},\ }\href {https://doi.org/10.1038/s41467-019-10988-2} {\bibfield  {journal} {\bibinfo  {journal} {Nature Communications}\ }\textbf {\bibinfo {volume} {10}},\ \bibinfo {pages} {3007} (\bibinfo {year} {2019})}\BibitemShut {NoStop}%
\bibitem [{\citenamefont {Tang}\ \emph {et~al.}(2021)\citenamefont {Tang}, \citenamefont {Shkolnikov}, \citenamefont {Barron}, \citenamefont {Grimsley}, \citenamefont {Mayhall}, \citenamefont {Barnes},\ and\ \citenamefont {Economou}}]{tang2021qubit}%
  \BibitemOpen
  \bibfield  {author} {\bibinfo {author} {\bibfnamefont {H.~L.}\ \bibnamefont {Tang}}, \bibinfo {author} {\bibfnamefont {V.~O.}\ \bibnamefont {Shkolnikov}}, \bibinfo {author} {\bibfnamefont {G.~S.}\ \bibnamefont {Barron}}, \bibinfo {author} {\bibfnamefont {H.~R.}\ \bibnamefont {Grimsley}}, \bibinfo {author} {\bibfnamefont {N.~J.}\ \bibnamefont {Mayhall}}, \bibinfo {author} {\bibfnamefont {E.}~\bibnamefont {Barnes}},\ and\ \bibinfo {author} {\bibfnamefont {S.~E.}\ \bibnamefont {Economou}},\ }\href {https://doi.org/10.1103/PRXQuantum.2.020310} {\bibfield  {journal} {\bibinfo  {journal} {PRX Quantum}\ }\textbf {\bibinfo {volume} {2}},\ \bibinfo {pages} {020310} (\bibinfo {year} {2021})}\BibitemShut {NoStop}%
\bibitem [{\citenamefont {Shkolnikov}\ \emph {et~al.}(2023)\citenamefont {Shkolnikov}, \citenamefont {Mayhall}, \citenamefont {Economou},\ and\ \citenamefont {Barnes}}]{Shkolnikov2023Symmetry}%
  \BibitemOpen
  \bibfield  {author} {\bibinfo {author} {\bibfnamefont {V.~O.}\ \bibnamefont {Shkolnikov}}, \bibinfo {author} {\bibfnamefont {N.~J.}\ \bibnamefont {Mayhall}}, \bibinfo {author} {\bibfnamefont {S.~E.}\ \bibnamefont {Economou}},\ and\ \bibinfo {author} {\bibfnamefont {E.}~\bibnamefont {Barnes}},\ }\href {https://doi.org/10.22331/q-2023-06-12-1040} {\bibfield  {journal} {\bibinfo  {journal} {Quantum}\ }\textbf {\bibinfo {volume} {7}},\ \bibinfo {pages} {1040} (\bibinfo {year} {2023})}\BibitemShut {NoStop}%
\bibitem [{\citenamefont {Liu}\ \emph {et~al.}(2021)\citenamefont {Liu}, \citenamefont {Li},\ and\ \citenamefont {Yang}}]{Liu2021AdaptiveRDMSolver}%
  \BibitemOpen
  \bibfield  {author} {\bibinfo {author} {\bibfnamefont {J.}~\bibnamefont {Liu}}, \bibinfo {author} {\bibfnamefont {Z.}~\bibnamefont {Li}},\ and\ \bibinfo {author} {\bibfnamefont {J.}~\bibnamefont {Yang}},\ }\href {https://doi.org/10.1063/5.0054822} {\bibfield  {journal} {\bibinfo  {journal} {The Journal of Chemical Physics}\ }\textbf {\bibinfo {volume} {154}},\ \bibinfo {pages} {244112} (\bibinfo {year} {2021})}\BibitemShut {NoStop}%
\bibitem [{\citenamefont {Sapova}\ and\ \citenamefont {Fedorov}(2022)}]{Sapova2022}%
  \BibitemOpen
  \bibfield  {author} {\bibinfo {author} {\bibfnamefont {M.~D.}\ \bibnamefont {Sapova}}\ and\ \bibinfo {author} {\bibfnamefont {A.~K.}\ \bibnamefont {Fedorov}},\ }\href {https://doi.org/10.1038/s42005-022-00982-4} {\bibfield  {journal} {\bibinfo  {journal} {Communications Physics}\ }\textbf {\bibinfo {volume} {5}},\ \bibinfo {pages} {199} (\bibinfo {year} {2022})}\BibitemShut {NoStop}%
\bibitem [{\citenamefont {Lan}\ and\ \citenamefont {Liang}(2022)}]{Lan2022AmplitudeReordering}%
  \BibitemOpen
  \bibfield  {author} {\bibinfo {author} {\bibfnamefont {Z.}~\bibnamefont {Lan}}\ and\ \bibinfo {author} {\bibfnamefont {W.}~\bibnamefont {Liang}},\ }\href {https://doi.org/10.1021/acs.jctc.2c00403} {\bibfield  {journal} {\bibinfo  {journal} {Journal of Chemical Theory and Computation}\ }\textbf {\bibinfo {volume} {18}},\ \bibinfo {pages} {5267} (\bibinfo {year} {2022})}\BibitemShut {NoStop}%
\bibitem [{\citenamefont {Anastasiou}\ \emph {et~al.}(2023)\citenamefont {Anastasiou}, \citenamefont {Mayhall}, \citenamefont {Barnes},\ and\ \citenamefont {Economou}}]{anastasiou2023measure}%
  \BibitemOpen
  \bibfield  {author} {\bibinfo {author} {\bibfnamefont {P.~G.}\ \bibnamefont {Anastasiou}}, \bibinfo {author} {\bibfnamefont {N.~J.}\ \bibnamefont {Mayhall}}, \bibinfo {author} {\bibfnamefont {E.}~\bibnamefont {Barnes}},\ and\ \bibinfo {author} {\bibfnamefont {S.~E.}\ \bibnamefont {Economou}},\ }\bibfield  {journal} {\bibinfo  {journal} {arXiv preprint}\ }\href {https://doi.org/10.48550/arXiv.2306.03227} {10.48550/arXiv.2306.03227} (\bibinfo {year} {2023})\BibitemShut {NoStop}%
\bibitem [{\citenamefont {Ikhtiarudin}\ \emph {et~al.}(2025)\citenamefont {Ikhtiarudin}, \citenamefont {Sunnardianto}, \citenamefont {Fathurrahman},\ and\ \citenamefont {Dipojono}}]{Ikhtiarudin2025}%
  \BibitemOpen
  \bibfield  {author} {\bibinfo {author} {\bibfnamefont {A.}~\bibnamefont {Ikhtiarudin}}, \bibinfo {author} {\bibfnamefont {G.~K.}\ \bibnamefont {Sunnardianto}}, \bibinfo {author} {\bibfnamefont {F.}~\bibnamefont {Fathurrahman}},\ and\ \bibinfo {author} {\bibfnamefont {H.~K.}\ \bibnamefont {Dipojono}},\ }\href {https://arxiv.org/abs/2507.16879} {\bibfield  {journal} {\bibinfo  {journal} {arXiv preprint arXiv:2507.16879}\ } (\bibinfo {year} {2025})}\BibitemShut {NoStop}%
\bibitem [{\citenamefont {Schr{\"o}der}\ \emph {et~al.}(2022)\citenamefont {Schr{\"o}der}, \citenamefont {Feldmann}, \citenamefont {Chancellor},\ and\ \citenamefont {Mintert}}]{AIMADAPTVQE2022}%
  \BibitemOpen
  \bibfield  {author} {\bibinfo {author} {\bibfnamefont {L.}~\bibnamefont {Schr{\"o}der}}, \bibinfo {author} {\bibfnamefont {P.}~\bibnamefont {Feldmann}}, \bibinfo {author} {\bibfnamefont {N.}~\bibnamefont {Chancellor}},\ and\ \bibinfo {author} {\bibfnamefont {F.}~\bibnamefont {Mintert}},\ }\href {https://arxiv.org/abs/2212.09719} {\bibfield  {journal} {\bibinfo  {journal} {arXiv preprint arXiv:2212.09719}\ } (\bibinfo {year} {2022})},\ \Eprint {https://arxiv.org/abs/2212.09719} {arXiv:2212.09719 [quant-ph]} \BibitemShut {NoStop}%
\bibitem [{\citenamefont {Van~Dyke}\ \emph {et~al.}(2024)\citenamefont {Van~Dyke}, \citenamefont {Shirali}, \citenamefont {Barron}, \citenamefont {Mayhall}, \citenamefont {Barnes},\ and\ \citenamefont {Economou}}]{vandyke2024operator}%
  \BibitemOpen
  \bibfield  {author} {\bibinfo {author} {\bibfnamefont {J.~S.}\ \bibnamefont {Van~Dyke}}, \bibinfo {author} {\bibfnamefont {K.}~\bibnamefont {Shirali}}, \bibinfo {author} {\bibfnamefont {G.~S.}\ \bibnamefont {Barron}}, \bibinfo {author} {\bibfnamefont {N.~J.}\ \bibnamefont {Mayhall}}, \bibinfo {author} {\bibfnamefont {E.}~\bibnamefont {Barnes}},\ and\ \bibinfo {author} {\bibfnamefont {S.~E.}\ \bibnamefont {Economou}},\ }\href {https://doi.org/10.1103/PhysRevResearch.6.L012030} {\bibfield  {journal} {\bibinfo  {journal} {Phys. Rev. Research}\ }\textbf {\bibinfo {volume} {6}},\ \bibinfo {pages} {L012030} (\bibinfo {year} {2024})}\BibitemShut {NoStop}%
\bibitem [{\citenamefont {Even-Dar}\ \emph {et~al.}(2006)\citenamefont {Even-Dar}, \citenamefont {Mannor},\ and\ \citenamefont {Mansour}}]{EvenDar2006Action}%
  \BibitemOpen
  \bibfield  {author} {\bibinfo {author} {\bibfnamefont {E.}~\bibnamefont {Even-Dar}}, \bibinfo {author} {\bibfnamefont {S.}~\bibnamefont {Mannor}},\ and\ \bibinfo {author} {\bibfnamefont {Y.}~\bibnamefont {Mansour}},\ }\href@noop {} {\bibfield  {journal} {\bibinfo  {journal} {Journal of Machine Learning Research}\ }\textbf {\bibinfo {volume} {7}},\ \bibinfo {pages} {1079} (\bibinfo {year} {2006})}\BibitemShut {NoStop}%
\bibitem [{\citenamefont {Audibert}\ \emph {et~al.}(2010)\citenamefont {Audibert}, \citenamefont {Bubeck},\ and\ \citenamefont {Munos}}]{Audibert2010BestArm}%
  \BibitemOpen
  \bibfield  {author} {\bibinfo {author} {\bibfnamefont {J.-Y.}\ \bibnamefont {Audibert}}, \bibinfo {author} {\bibfnamefont {S.}~\bibnamefont {Bubeck}},\ and\ \bibinfo {author} {\bibfnamefont {R.}~\bibnamefont {Munos}},\ }in\ \href@noop {} {\emph {\bibinfo {booktitle} {COLT}}}\ (\bibinfo {year} {2010})\BibitemShut {NoStop}%
\bibitem [{\citenamefont {Bubeck}\ and\ \citenamefont {Cesa-Bianchi}(2012)}]{Bubeck2012Bandits}%
  \BibitemOpen
  \bibfield  {author} {\bibinfo {author} {\bibfnamefont {S.}~\bibnamefont {Bubeck}}\ and\ \bibinfo {author} {\bibfnamefont {N.}~\bibnamefont {Cesa-Bianchi}},\ }\href {https://doi.org/10.1561/2200000024} {\bibfield  {journal} {\bibinfo  {journal} {Foundations and Trends in Machine Learning}\ }\textbf {\bibinfo {volume} {5}},\ \bibinfo {pages} {1} (\bibinfo {year} {2012})}\BibitemShut {NoStop}%
\bibitem [{\citenamefont {Patel}\ \emph {et~al.}(2025)\citenamefont {Patel}, \citenamefont {Jayakumar}, \citenamefont {Yen},\ and\ \citenamefont {Izmaylov}}]{patel2025quantum}%
  \BibitemOpen
  \bibfield  {author} {\bibinfo {author} {\bibfnamefont {S.}~\bibnamefont {Patel}}, \bibinfo {author} {\bibfnamefont {P.}~\bibnamefont {Jayakumar}}, \bibinfo {author} {\bibfnamefont {T.-C.}\ \bibnamefont {Yen}},\ and\ \bibinfo {author} {\bibfnamefont {A.~F.}\ \bibnamefont {Izmaylov}},\ }\href@noop {} {\bibfield  {journal} {\bibinfo  {journal} {Chemical Reviews}\ }\textbf {\bibinfo {volume} {125}},\ \bibinfo {pages} {7490} (\bibinfo {year} {2025})}\BibitemShut {NoStop}%
\bibitem [{\citenamefont {Yen}\ \emph {et~al.}(2023)\citenamefont {Yen}, \citenamefont {Ganeshram},\ and\ \citenamefont {Izmaylov}}]{yen2023deterministic}%
  \BibitemOpen
  \bibfield  {author} {\bibinfo {author} {\bibfnamefont {T.-C.}\ \bibnamefont {Yen}}, \bibinfo {author} {\bibfnamefont {A.}~\bibnamefont {Ganeshram}},\ and\ \bibinfo {author} {\bibfnamefont {A.~F.}\ \bibnamefont {Izmaylov}},\ }\href@noop {} {\bibfield  {journal} {\bibinfo  {journal} {npj Quantum Inf}\ }\textbf {\bibinfo {volume} {9}} (\bibinfo {year} {2023})}\BibitemShut {NoStop}%
\bibitem [{\citenamefont {Yordanov}\ \emph {et~al.}(2021)\citenamefont {Yordanov}, \citenamefont {Armaos}, \citenamefont {Barnes},\ and\ \citenamefont {Arvidsson-Shukur}}]{yordanov2021qubit}%
  \BibitemOpen
  \bibfield  {author} {\bibinfo {author} {\bibfnamefont {Y.~S.}\ \bibnamefont {Yordanov}}, \bibinfo {author} {\bibfnamefont {V.}~\bibnamefont {Armaos}}, \bibinfo {author} {\bibfnamefont {C.~H.~W.}\ \bibnamefont {Barnes}},\ and\ \bibinfo {author} {\bibfnamefont {D.~R.~M.}\ \bibnamefont {Arvidsson-Shukur}},\ }\href {https://doi.org/10.1038/s42005-021-00730-0} {\bibfield  {journal} {\bibinfo  {journal} {Communications Physics}\ }\textbf {\bibinfo {volume} {4}},\ \bibinfo {pages} {228} (\bibinfo {year} {2021})}\BibitemShut {NoStop}%
\bibitem [{\citenamefont {Garivier}\ and\ \citenamefont {Kaufmann}(2016)}]{pmlr-v49-garivier16a}%
  \BibitemOpen
  \bibfield  {author} {\bibinfo {author} {\bibfnamefont {A.}~\bibnamefont {Garivier}}\ and\ \bibinfo {author} {\bibfnamefont {E.}~\bibnamefont {Kaufmann}},\ }in\ \href {https://proceedings.mlr.press/v49/garivier16a.html} {\emph {\bibinfo {booktitle} {29th Annual Conference on Learning Theory}}},\ \bibinfo {series} {Proceedings of Machine Learning Research}, Vol.~\bibinfo {volume} {49},\ \bibinfo {editor} {edited by\ \bibinfo {editor} {\bibfnamefont {V.}~\bibnamefont {Feldman}}, \bibinfo {editor} {\bibfnamefont {A.}~\bibnamefont {Rakhlin}},\ and\ \bibinfo {editor} {\bibfnamefont {O.}~\bibnamefont {Shamir}}}\ (\bibinfo  {publisher} {PMLR},\ \bibinfo {address} {Columbia University, New York, New York, USA},\ \bibinfo {year} {2016})\ pp.\ \bibinfo {pages} {998--1027}\BibitemShut {NoStop}%
\bibitem [{\citenamefont {Russo}(2020)}]{russo2020simple}%
  \BibitemOpen
  \bibfield  {author} {\bibinfo {author} {\bibfnamefont {D.}~\bibnamefont {Russo}},\ }\href {https://doi.org/10.1287/opre.2019.1911} {\bibfield  {journal} {\bibinfo  {journal} {Operations Research}\ }\textbf {\bibinfo {volume} {68}},\ \bibinfo {pages} {1625} (\bibinfo {year} {2020})}\BibitemShut {NoStop}%
\end{thebibliography}%

\end{document}